\documentclass[aps,preprint,superscriptaddress,nofootinbib,showpacs]{revtex4}
\usepackage[dvipdfmx]{graphicx}
\usepackage{bm,latexsym,amsmath,amssymb,amsfonts,mathrsfs}
\usepackage{color}
\input{colordvi.tex}
\usepackage[dvipdfmx]{hyperref}
\usepackage{url}
\hypersetup{
    colorlinks=true,
    citecolor=cyan,
}
\newcommand*{\D}{{\rm d}}
\newcommand*{\mpl}{M_{\rm Pl}}
\begin{document}

\title{Removing Ostrogradski's ghost from cosmological perturbations in $f(R,R_{\mu\nu}^2,C_{\mu\nu\rho\sigma}^2)$ gravity}

\author{Yuji~Akita}
\affiliation{Department of Physics, Rikkyo University, Toshima, Tokyo 175-8501, Japan
}

\author{Tsutomu~Kobayashi}
\affiliation{Department of Physics, Rikkyo University, Toshima, Tokyo 175-8501, Japan
}

\begin{abstract}
Recently it was argued that gravity with the squire of the Ricci tensor
can be stabilized by adding constraints to the theory.
This was so far demonstrated for fluctuations on the Minkowski/de Sitter background.
We show that the same scheme works equally well for removing Ostrogradski's ghost from fluctuations on
a cosmological background in generic $f(R,R_{\mu\nu}^2,C_{\mu\nu\rho\sigma}^2)$-type theories of gravity.
We also derive the general formula for the spectrum of primordial tensor perturbations
from the stabilized theory.
\end{abstract}

\pacs{
04.50.Kd  
}
\preprint{RUP-15-15}
\maketitle
\section{Introduction}

Physical laws describing the time evolution are written in the form of
differential equations up to second order in time.
This is indeed the case for classical mechanics, Maxwell's theory of electromagnetism,
and general relativity. Then, what if one has the evolution equations of higher order in time?
The answer is tragic; one will encounter
{\em Ostrogradski's ghost} that is the generic instability
in non-degenerate higher derivative theories~\cite{Ostrogradski, Woodard:2006nt, Woodard:2015zca}.
Ostrogradski's instability can be illustrated by the following simple example~\cite{Chen:2012au}: 
	\begin{eqnarray}
	S=\int\D t
	 \left[\frac{1}{2}\ddot{q}^2-V(q)\right].
	\end{eqnarray}
This action yields the fourth-order equation of motion for $q$.
Defining the canonical coordinates $Q_1:=q$, $Q_2:=\dot q$ and their conjugate momenta $P_1$, $P_2$,
one obtains the Hamiltonian,
	\begin{eqnarray}
	H=P_1Q_2+\frac{P_2^2}{2}+V(Q_1).
	\end{eqnarray}
This Hamiltonian linearly depends on $P_1$ and hence
is not bounded from below, signaling the instability.

General relativity, a second-order theory for the metric, is a healthy theory from this viewpoint.
Nevertheless, it is sufficiently reasonable to consider gravitational theories beyond general relativity.
Higher powers of the curvature tensors
such as $R^2$ and $R_{\mu\nu}R^{\mu\nu}$
are expected to be the low-energy manifestation of
quantum gravity. In recent years
phenomenological modification of general relativity has been studied extensively
in order to account for the present accelerated expansion of the Universe,
which involves an arbitrary function of the curvature tensors or a
non-minimal coupling to the scalar field.
Due to the higher derivative nature, those theories are often plagued by Ostrogradski's instability.

There are several ways to evade this instability issue.
From an effective field theory point of view, if one takes the UV cutoff below the
scale at which the ghost emerges then the instability is not necessarily problematic
and thus could be circumvented.
More directly, one would consider the class of theories that violates the assumptions of Ostrogradski's theorem.
The $R^2$ model~\cite{r2} and its $f(R)$ generalization~\cite{Sotiriou:2008rp,DeFelice:2010aj}
are degenerate, and hence are free of ghosts.
The Galileons~\cite{Nicolis:2008in,Deffayet:2009wt,Deffayet:2009mn} and
the Horndeski family of scalar-tensor theories~\cite{Horndeski:1974wa,Deffayet:2011gz,Kobayashi:2011nu}
have manifestly second-order field equations
despite the higher derivative nature of the Lagrangian,
and therefore we do not need to care about Ostrogradski's ghost.
(In fact, it is well known that $f(R)$ theories can be recast in a
scalar-tensor theory that is in a subclass of the Horndeski theory.)
Recently, an approach in this direction has been pushed forward and
healthy theories beyond Horndeski have been
developed~\cite{Gleyzes:2014dya,Gao:2014soa,Gleyzes:2014qga,Lin:2014jga,Gao:2014fra,Deffayet:2015qwa,Zumalacarregui:2013pma}.

Yet another way of getting around the ghost is adding constraints
to the theory to reduce the dimensionality of the phase space,
as proposed in Refs.~\cite{Chen:2012au,Chen:2013aha}.
In Ref.~\cite{Chen:2013aha}, the theory described by~\cite{Stelle:1976gc}
	\begin{eqnarray}
	S=\frac{\mpl^2}{2}\int \D^4x\sqrt{-g}\left(R-2\Lambda+\alpha R^2+\beta R_{\mu\nu}R^{\mu\nu}\right)
	\end{eqnarray}
was examined. Although this theory contains unstable modes as it is,
it can be made stable by the appropriate addition of constraints at least
at the level of the quadratic action for fluctuations.
This was demonstrated for the fluctuations on the Minkowski and de Sitter backgrounds~\cite{Chen:2013aha}.

The purpose of the present paper is to extend the work of~\cite{Chen:2013aha}.
We consider the more general action of the form
	\begin{eqnarray}
	S=\int\D^4x\sqrt{-g}f(R, R_{\mu\nu}R^{\mu\nu}, C_{\mu\nu\rho\sigma}C^{\mu\nu\rho\sigma}),\label{action}
	\end{eqnarray}
where $f$ is an arbitrary function, and stabilize the quadratic action
for the fluctuations on the Friedmann-Lma\^{i}tre-Robertson-Walker (FLRW) background.
Note that since the Weyl tensor $C_{\mu\nu\rho\sigma}$ can be expressed in terms of
the Riemann tensor, the Ricci tensor, and the Ricci scalar, the above theory is
nothing but $f(R, R_{\mu\nu}^2, R_{\mu\nu\rho\sigma}^2)$ gravity.
This class of gravitational theories has been explored
extensively~\cite{Hindawi:1995an,Hindawi:1995cu,Carroll:2004de,Nunez:2004ts,Chiba:2005nz,Calcagni:2005im,Navarro:2005da,DeFelice:2006pg}.
As an application,
we derive the general formula for
the primordial power spectrum of the tensor modes from the stabilized action.

The paper is organized as follows.
In the next section we give a quick review on
the instabilities in $f(R, R_{\mu\nu}^2, C_{\mu\nu\rho\sigma}^2)$ gravity.
We then stabilize the quadratic action for cosmological perturbations
by adding constraints in Sec.~III.
In Sec.~IV, we evaluate the amplitude of primordial tensor modes in the stabilized theory.
Section~V is devoted to discussion and conclusions.

\section{Instabilities in $f(R,R_{\mu\nu}^2,C_{\mu\nu\rho\lambda}^2)$ gravity}

We start with reviewing how ghost instabilities appear
in the theory
described by the action~(\ref{action}).
The background we consider is given by the FLRW metric,
$\D s^2=a^2(\eta)\left(-\D\eta^2+\delta_{ij}\D x^i\D x^j\right)$.
Let us use the notation
\begin{eqnarray}
r_1:=R,\quad
r_2:=R_{\mu\nu}R^{\mu\nu},\quad
r_3:=C_{\mu\nu\rho\sigma}C^{\mu\nu\rho\sigma},
\end{eqnarray}
and write $f_0:=f$,
$f_{i}:=\partial f/\partial r_i$, $f_{ij}:=\partial^2 f/\partial r_i\partial r_j$,
and $f_{ijk}:=\partial^3 f/\partial r_i\partial r_j\partial r_k$
evaluated at the background, {\em i.e.},
$r_1=6\left({\cal H}^2+{\cal H}'\right)/a^2$,
$r_2=12\left[{\cal H}^4+{\cal H}^2{\cal H}'+({\cal H}')^2\right]/a^4$, and $r_3=0$,
where ${\cal H}:=a'/a$ and the prime denotes differentiation with respect to the conformal time $\eta$.

The background equation can be derived simply by substituting
the metric $\D s^2=-N^2(\eta)\D\eta^2+a^2(\eta)\delta_{ij}\D x^i\D x^j$ to the
action~(\ref{action}), varying it with respect to $N$ and $a$,
and then setting $N=a$.
Variation with respect to $N$ gives
\begin{eqnarray}
0&=&
	a^4 f_0
	-6 a^2 \mathcal{H}' f_1
	-12\left[\left(2 f_2+3 f_{11}\right) \mathcal{H}\left(2\mathcal{H}^3-\mathcal{H}'' \right)+f_2 \mathcal{H}' \left(\mathcal{H}^2+2 \mathcal{H}'\right)\right]
\nonumber\\&&
	-\frac{144}{a^2} f_{12} \mathcal{H} \left[3 \mathcal{H}^5-\mathcal{H}^2 \mathcal{H}''+\mathcal{H} \left(\mathcal{H}'\right)^2+2 \mathcal{H}^3 \mathcal{H}'-2 \mathcal{H}' \mathcal{H}''\right]
\nonumber\\&&
	-\frac{144}{a^4} f_{22} \mathcal{H} \left(\mathcal{H}^2+2 \mathcal{H}'\right) \left[4 \mathcal{H}^5-\mathcal{H}^2 \mathcal{H}''+2 \mathcal{H} \left(\mathcal{H}'\right)^2-2 \mathcal{H}' \mathcal{H}''\right].
\end{eqnarray}
This equation will be used in the following calculations
to eliminate or rearrange some coefficients in the quadratic actions for fluctuations.
Variation with respect to $a$ gives another cumbersome equation
which is not used in the present paper.

\subsection{Tensor perturbations}

Let us first look at Ostrogradski's instability of tensor perturbations.
It is convenient to parametrize the tensor perturbations as
\begin{eqnarray}
\D s^2=a^2\left[-\D\eta^2+\left( \delta_{ij}+h_{ij}+\frac{1}{2}h_{ik}h_{kj}\right)\D x^i\D x^j\right],
\end{eqnarray}
as this definition yields $\sqrt{-g}=a^4$ at quadratic order in fluctuations.
Substituting this metric to the action~(\ref{action}), we obtain
the quadratic action for the tensor perturbations:
\begin{align}
S=\int\D^4x & \biggl\{
	a^2 f_1\left(\frac{1}{4}h_{ij}'^2+\frac{1}{4}h_{ij}\partial^2h_{ij}\right)
	+\frac{f_2{\cal H}}{2}\left(h_{ij}'^2-h_{ij}\partial^2h_{ij}\right)'
	\nonumber\\&
	+f_2\left[
	\frac{1}{4}h_{ij}''^2-\frac{1}{2}h_{ij}''\partial^2h_{ij}
	+\frac{1}{4}\left(\partial^2 h_{ij}\right)^2
	+\left({\cal H}^2+\frac{3}{2}{\cal H}'\right)h_{ij}'^2+\left({\cal H}^2+\frac{{\cal H}'}{2}\right)
	h_{ij}\partial^2h_{ij}
	\right]
	\notag\\&
	+f_3 \left[
	\frac{1}{2}h_{ij}''^2+ h_{ij}''\partial^2h_{ij}+2h_{ij}'\partial^2 h_{ij}'
	+\frac{1}{2}\left(\partial^2 h_{ij}\right)^2
	\right]
	\biggr\}.
\end{align}
This can be simplified by integration by parts as
\begin{eqnarray}\label{action: t}
S=\int\D^4x\left\{
	\frac{a^2}{8}\left(Ah_{ij}'^2+Ch_{ij}\partial^2h_{ij}\right)
	+\frac{\beta}{8}
	\left[h_{ij}''^2+2h_{ij}'\partial^2h_{ij}'+\left(\partial^2h_{ij}\right)^2\right]
	\right\},
\end{eqnarray}
where we defined
\begin{eqnarray}
\beta :=2f_2+4f_3,
\end{eqnarray}
and
\begin{eqnarray}
A&:=&2f_1+\frac{8f_2}{a^2}\left({\cal H}^2+{\cal H}'\right)-4f_2'\frac{{\cal H}}{a^2},
\\
C&:=&2f_1+\frac{8f_2}{a^2}\left({\cal H}^2+{\cal H}'\right)+\frac{4f_2'}{a^2}{\cal H}
-\frac{2}{a^2}\left(f_2''-2f_3''\right).
\end{eqnarray}
Throughout the paper we assume that $\beta=\beta(\eta)$ never vanishes.\footnote{If $f$
depends on the curvature invariants through the Gauss-Bonnet combination,
${\cal G}:=R_{\mu\nu\rho\lambda}^2-4R_{\mu\nu}^2+R^2$, {\em i.e.}, $f=f(R,{\cal G})$,
we identically have $\beta =0$. Therefore, we do not consider the $f(R,{\cal G})$ class of theories in this paper.
Linear cosmological perturbations are healthy in $f(R,{\cal G})$ gravity~\cite{DeFelice:2009ak},
though it was found in the end that
ghost degrees of freedom cannot be avoided on less symmetric backgrounds~\cite{Felice-Tanaka}.}
The resultant action turns out to be of the same form as
that on the Minkowski/de Sitter background in the $\alpha R^2+\beta R_{\mu\nu}R^{\mu\nu}$ theory~\cite{Chen:2013aha},
though in the present case the coefficients $A$, $C$, and $\beta$ are time-dependent in general.
The quadratic action~(\ref{action: t}) contains higher derivative terms,
and hence
Ostrogradski's instability appears as expected.

We confirm the presence of Ostrogradski's instability in a more rigorous way by the Hamiltonian analysis.
Defining the canonical coordinates as
\begin{eqnarray}
h_{ij}\equiv h_{ij}&\longleftrightarrow&
	\pi^{ij}=
	\frac{1}{4}\left[a^2Ah_{ij}'-\beta\left(h_{ij}'''-2\partial^2h_{ij}'\right)-\beta'h_{ij}''\right]~,
\\
q_{ij}\equiv h_{ij}' &\longleftrightarrow&
	p^{ij}=\frac{\beta}{4}h_{ij}''
	~,
\end{eqnarray}
we obtain the Hamiltonian
\begin{eqnarray}
H=\int\D^3x\left[
	\pi^{ij}q_{ij}+\frac{2}{\beta} p_{ij}^2
	-\frac{1}{8}h_{ij}\left(a^2C\partial^2+\beta\partial^2\partial^2\right)h_{ij}
	-\frac{1}{8}q_{ij}\left(a^2A+2\beta\partial^2\right)q_{ij}
	\right].
\end{eqnarray}
This Hamiltonian linearly depends on $\pi_{ij}$ as can be seen in the first term,
implying that the Hamiltonian is not bounded from below.
Thus, we see that the higher derivative nature of the action~(\ref{action: t})
gives rise to the ghost instability.

\subsection{Vector perturbations}

The vector sector of the metric perturbations is given by
\begin{eqnarray}
\D s^2=a^2\left[-\left(1-B_iB^i\right)
\D\eta^2+2B_i\D x^i\D\eta+\delta_{ij}\D x^i\D x^j\right],
\end{eqnarray}
with $\partial_iB_i =0$.
Here we added $B_iB^i$ in the $(00)$ component so that $\sqrt{-g}=a^4$.
Substituting the metric to the action~(\ref{action}), we obtain
the quadratic action for the vector perturbations:
\begin{eqnarray}
S=\int\D^4x\left[
	\frac{\beta}{4}\left(\hat v_{ij}'^2+\hat v_{ij}\partial^2\hat v_{ij}\right)
	+\frac{a^2A}{4} \hat v_{ij}^2
	\right]~,
\end{eqnarray}
where $\beta$ and $A$ are the same as
the corresponding quantities defined in the previous subsection, and $\hat v_{ij}:=\partial_iB_j$.
We thus see that in the theory~(\ref{action}) the vector perturbations are dynamical in general.
However, the vector sector is free of any instabilities provided that $\beta > 0$ and $A<0$.
Note that here again the quadratic action takes the same form as
that on the Minkowski/de Sitter background in the $\alpha R^2+\beta R_{\mu\nu}R^{\mu\nu}$ theory~\cite{Chen:2013aha},
but with the time-dependent coefficients.

In terms of the canonical momentum conjugate $\hat\pi^{ij} = (\beta/2)\hat v^{ij}$,
the Hamiltonian of the vector sector is written as
\begin{eqnarray}
H=\int\D^3x\left(
\frac{\hat\pi_{ij}\hat\pi^{ij}}{\beta}
-\frac{\beta}{4}\hat v_{ij}\partial^2\hat v_{ij}-\frac{a^2A}{4} \hat v_{ij}^2\right)~.
\end{eqnarray}
This Hamiltonian is bounded from below
and therefore the vector sector is stable
for $\beta >0$ and $A<0$.

\subsection{Scalar perturbations}

To simplify the manipulation we fix the gauge inside the action.
It is probably the most suitable to take the flat gauge, and
it is indeed possible to do so at the action level~\cite{Lagos:2013aua}.
The metric can thus be written as
\begin{eqnarray}
\D s^2=a^2 \left[ -(1+2\Phi )\D \eta^2+2\partial_i B\D \eta\D x^i+\delta_{ij}\D x^i\D x^j \right].
\end{eqnarray}
The quadratic action for the scalar perturbations is given by
\begin{eqnarray}
S&=&\int\D^4x\bigl[b_0\left(\Phi^\prime\right)^2+
	b_1\left(\partial^2\Phi+{\cal B}^\prime\right)^2
	+b_2\Phi^\prime{\cal B}^\prime
	+b_3\Phi{\cal B}^\prime
\nonumber\\&& \qquad \quad
	+c_1\Phi^2+c_2\Phi\partial^2\Phi+c_3\Phi{\cal B}+c_4{\cal B}^2+c_5{\cal B}\partial^2\Phi
	\bigr],\label{scalar-action}
\end{eqnarray}
where $\mathcal{B}:=\partial^2B$.
The coefficients in the above action are
\begin{eqnarray}
b_0&:=&6\mathcal{H}^2\mathcal{I}~,
\\
b_1&:=&\frac{2}{3}\mathcal{I}+\frac{\beta}{3}~,
\\
b_2&:=&4\mathcal{H}\mathcal{I}~,
\\
b_3&:=&4\left(2\mathcal{H}'-\mathcal{H}^2\right)\mathcal{I}~,
\end{eqnarray}
where
\begin{eqnarray}
\mathcal{I}&:=&
	2f_2+3f_{11}+\frac{12}{a^2}\left(\mathcal{H}^2+2\mathcal{H}'\right)f_{12}
	+\frac{12}{a^4}\left(\mathcal{H}^2+2\mathcal{H}'\right)^2f_{22}~.
\end{eqnarray}
The concrete expressions for $c_i$ ($i=$1--5) are lengthy and are summarized in Appendix~\ref{appcf}.

It is not obvious from~(\ref{scalar-action})
whether or not the scalar sector contains unstable degrees of freedom.
Let us therefore take a careful look at the Hamiltonian for the scalar perturbations.
We choose to use $(\Phi, {\cal B})$ as canonical coordinates, and then
the corresponding canonical momenta are given respectively by
\begin{eqnarray}
\pi_\Phi&=&2b_0\Phi'+b_2\mathcal{B}',
\\
\pi_\mathcal{B}&=&2b_1\left(\partial^2\Phi+\mathcal{B}'\right)+b_2\Phi'+b_3\Phi.
\end{eqnarray}
The Hamiltonian is
\begin{eqnarray}
H&=&\int\D^3x \biggl[
\frac{3}{4\beta}\left(\pi_{\cal B}-\frac{1}{3}\frac{\pi_{\Phi}}{{\cal H}}-b_3\Phi-2b_1\partial^2\Phi \right)^2
+\frac{\pi_{\Phi}^2}{24{\cal H}^2{\cal I}}
-\frac{2{\cal I}+\beta }{3} (\partial^2\Phi)^2
\nonumber\\&& \qquad
 -c_1\Phi^2-c_2\Phi\partial^2\Phi-c_3\Phi{\cal B}-c_4{\cal B}^2-c_5{\cal B}\partial^2\Phi
\biggr].
\end{eqnarray}
To see the instabilities it is more convenient to perform
a canonical transformation,
\begin{eqnarray}
\widetilde\pi_{\Phi}&=&\pi_{\Phi},
\\
\widetilde \pi_{\cal B}&=&\pi_{\cal B}-\frac{\pi_\Phi}{3{\cal H}}
+\frac{1}{3{\cal H}}\left(b_3{\cal B}+2b_1\partial^2{\cal B} \right),
\\
\widetilde\Phi&=&\Phi+\frac{{\cal B}}{3{\cal H}},
\\
\widetilde{\cal B}&=&{\cal B}.
\end{eqnarray}
The generating function $F$ of this transformation
is given by
\begin{eqnarray}
F
=\int\D^3x\left[\widetilde\pi_\Phi\left(\Phi+\frac{{\cal B}}{3{\cal H}}\right)+\widetilde\pi_{\cal B}{\cal B}
-\frac{1}{6{\cal H}}\left(b_3{\cal B}^2+2b_1{\cal B}\partial^2{\cal B}\right)\right].
\end{eqnarray}
The above canonical transformation turns the Hamiltonian into
\begin{eqnarray}
\widetilde H=\int\D^3x\Biggl\{
\frac{3}{4\beta}\left(\widetilde\pi_{\cal B}-\widetilde\Phi b_3-2b_1\partial^2\widetilde\Phi\right)^2
+\frac{\widetilde\pi_\Phi^2}{24{\cal H}^2{\cal I}}-\frac{2{\cal I}+\beta}{3}
\left[\partial^2\left(\widetilde\Phi-\frac{\widetilde{\cal B}}{3{\cal H}}\right)\right]^2+\cdots
\Biggr\},
\end{eqnarray}
where we have written only the terms that are relevant at high momenta.
This Hamiltonian clearly shows how instabilities arise:
it is required for
stable kinetic terms that $\beta>0$ and ${\cal I}>0$, but then
the gradient instability occurs at high momenta, as seen in the coefficient of the $(\partial^2\widetilde{\cal B})^2$ term.

\section{Stabilization of $f(R,R_{\mu\nu}^2,C_{\mu\nu\rho\sigma}^2)$ gravity}

In the previous section we have reviewed instabilities of cosmological perturbations
in $f(R,R_{\mu\nu}^2, C_{\mu\nu\rho\sigma}^2)$ gravity.
For all types of perturbations, it is found that the basic structure of the
quadratic actions
is very similar to that on the Minkowski/de Sitter background in $\alpha R^2+\beta R_{\mu\nu}R^{\mu\nu}$
gravity~\cite{Chen:2013aha}.
This suggests that one can
stabilize the theory,
in the same way as in Ref.~\cite{Chen:2013aha},
by imposing suitable constraints while maintaining the
renormalization properties.
In this section, we demonstrate that this is indeed true.

\subsection{Tensor perturbations}

Following Ref.~\cite{Chen:2013aha}, we
introduce a auxiliary tensor field $\lambda_{ij}$ into the action~(\ref{action: t}):
\begin{eqnarray}\label{action: t with lambda}
S&=&\int\D^4x\left\{
	\frac{a^2}{8}\left(Ah_{ij}'^2+C h_{ij}\partial^2h_{ij}\right)
	\right. \nonumber\\&& \left.\qquad\qquad
	+\frac{\beta}{8}
	\left[\left(h_{ij}''-\lambda_{ij}\right)^2+2h_{ij}'\partial^2h_{ij}'+\left(\partial^2h_{ij}\right)^2\right]
	+\frac{\beta}{2}\lambda_{ij}\partial^2h_{ij}
	\right\}.
\end{eqnarray}
Variation with respect to $\lambda_{ij}$ gives the constraint
\begin{eqnarray}
\lambda_{ij}-h_{ij}''+2\partial^2h_{ij}=0~.
\end{eqnarray}
Substituting this constraint back into the original action~(\ref{action: t with lambda}), we arrive at
\begin{eqnarray}
S=\frac{1}{8}\int\D^4x\left\{
	h_{ij}'\left(a^2A-2\beta\partial^2\right)h_{ij}'
	+h_{ij}\left[\left(a^2C+2\beta''\right)\partial^2-3\beta\partial^2\partial^2\right]h_{ij}\right\}.
	\label{2nd-ac-tens-cost}
\end{eqnarray}
The dangerous second time derivative $h_{ij}''$ can thus be removed from the action
while retaining higher spatial derivatives acting on $h_{ij}$ and $h_{ij}'$.
(The spirit here is the same as that of Ho\v{r}ava gravity which is power-counting renormalizable~\cite{Horava}.)
The reduced action implies that
the tensor sector becomes free of Ostrogradski's instability.

To confirm the stability let us construct the Hamiltonian of the tensor sector.
The canonical coordinates we choose are
\begin{eqnarray}
h_{ij}\equiv h_{ij}&\longleftrightarrow&
	\pi^{ij}=
	\frac{1}{4}\left[a^2Ah_{ij}'-\beta\left(h_{ij}'''-\lambda_{ij}'-2\partial^2h_{ij}'\right)-\beta'(h_{ij}''-\lambda_{ij})\right],
\\
q_{ij}\equiv h_{ij}' &\longleftrightarrow&
	p^{ij}=\frac{\beta}{4}\left(h_{ij}''-\lambda_{ij}\right)
	,
\\
\lambda_{ij}\equiv \lambda_{ij} &\longleftrightarrow& \pi_\lambda^{ij}=0.
\end{eqnarray}
The Hamiltonian is then given by
\begin{eqnarray}
H&=&\int\D^3x\biggl[
	\pi^{ij}q_{ij}+\frac{2}{\beta} p_{ij}^2+\lambda_{ij}\left(p_{ij}-\frac{\beta}{2}\partial^2h_{ij}\right)
\nonumber\\&& \qquad \qquad
	-\frac{1}{8}h_{ij}\left(a^2C\partial^2+\beta\partial^2\partial^2\right)h_{ij}
	-\frac{1}{8}q_{ij}\left(a^2A+2\beta\partial^2\right)q_{ij}
	\biggr].
\end{eqnarray}
The primary constraint is $\pi_\lambda^{ij}=0$.
The consistency of the constraints
generates the following set of secondary constraints:
\begin{eqnarray}
&&
p_{ij}-\frac{\beta}{2}\partial^2h_{ij}\approx 0,
\\
&&
\pi_{ij}-\frac{1}{4}a^2A q_{ij}\approx 0,
\\
&&
\left(a^2A-2\beta \partial^2\right)\lambda_{ij}+\frac{4}{\beta}a^2Ap_{ij}
-\left(a^2C\partial^2+\beta\partial^2\partial^2\right)h_{ij}\approx 0,
\end{eqnarray}
where $\approx$ stands for weak equality.
These are the second class constraints.
We use these constraints to eliminate $(\lambda_{ij},\pi_\lambda^{ij})$ and $(q_{ij},p_{ij})$,
and obtain the reduced Hamiltonian,
\begin{eqnarray}
H_R=\int\D^3x \left[
	\frac{2}{a^4A^2}\pi_{ij}\left(a^2A-2\beta\partial^2\right)\pi_{ij}
	+\frac{1}{8}h_{ij}\left(-a^2C\partial^2+3\beta\partial^2\partial^2\right)h_{ij}
	\right],
\end{eqnarray}
which is positive definite if
\begin{eqnarray}
\beta(\eta) >0,\quad A(\eta)>0, \quad C(\eta)>0. \label{tensor-st}
\end{eqnarray}
This is the sufficient conditions for the absence of instabilities.
Note, however, that if $C$ becomes negative for a sufficiently
short period then the tensor sector is still stable because
only low momentum modes develop instabilities whose time scales are bounded from below.

Since each tensor variable has two independent components,
the original theory~(\ref{action: t}) contains eight degrees of freedom in phase space.
We then add four second class constraints, leaving four degrees of freedom in phase space
in the constrained theory. The ghost modes in the tensor sector can thus be removed.

\subsection{Vector perturbations}

The vector perturbations are not harmful in itself; as we have seen in the previous section,
the vector modes are stable for $\beta>0$ and $A<0$.
However, this is incompatible with the stability conditions for the tensor perturbations~(\ref{tensor-st}).
For this reason, we are going to remove the vector modes from the theory.

Introducing a auxiliary vector field $\lambda_{i}$,
we consider the modified quadratic action
\begin{eqnarray}
S=\int\D^4x\left\{
	\frac{\beta}{4}\left[\left(\hat v_{ij}'-\hat\lambda_{ij}\right)^2
	+\hat v_{ij}\partial^2\hat v_{ij}\right]+\frac{a^2A}{4} \hat v_{ij}^2
	\right\},
\end{eqnarray}
with $\hat\lambda_{ij}:=\partial_i\lambda_j$.
The canonical coordinates are chosen to be
\begin{eqnarray}
\hat v_{ij}\equiv \hat v_{ij} &\longleftrightarrow&
\hat\pi^{ij}= \frac{\beta}{2}\left(\hat v^{ij}-\hat\lambda^{ij}\right),
\\
\hat \lambda_{ij}\equiv \hat \lambda_{ij} &\longleftrightarrow&
\hat \pi_\lambda^{ij}=0,
\end{eqnarray}
and the Hamiltonian is
\begin{eqnarray}
H=\int\D^3x\left(
	\frac{\hat\pi_{ij}^2}{\beta}+\hat \pi_{ij}\hat \lambda_{ij}
	-\frac{\beta}{4}\hat v_{ij}\partial^2\hat v_{ij}-\frac{a^2A}{4} \hat v_{ij}^2
	\right).
\end{eqnarray}
We can derive secondary constraints from the primary constraint $\hat \pi_\lambda^{ij}=0$ as
\begin{eqnarray}
&&
\hat\pi^{ij}\approx 0,
\\
&&
\left(a^2A+\beta\partial^2\right)\hat v_{ij}\approx 0,
\\
&&
\left(a^2A+\beta\partial^2 \right)\left(\hat\lambda_{ij}+\frac{2}{\beta}\hat \pi_{ij}\right) \approx 0.
\end{eqnarray}
Substituting these constraints back to the Hamiltonian,
we see that the reduced Hamiltonian vanishes, indicating that there are no vector degrees of freedom.

\subsection{Scalar perturbations}

Along the same line as the stabilization procedure on the de Sitter background~\cite{Chen:2013aha},
we introduce a auxiliary scalar field $\lambda$ to modify
the quadratic action as
\begin{eqnarray}
S&=&\int\D^4x\bigl[b_0(\Phi')^2+
	b_1\left(\partial^2\Phi+{\cal B}^\prime-\lambda\right)^2
	+b_2\Phi^\prime({\cal B}^\prime-\lambda)
	+b_3\Phi({\cal B}^\prime-\lambda)
\nonumber\\&& \qquad \quad
	+c_1\Phi^2+c_2\Phi\partial^2\Phi+c_3\Phi{\cal B}+c_4{\cal B}^2	+c_5{\cal B}\partial^2\Phi
	\bigr].
\end{eqnarray}
The canonical momenta are now given by
\begin{eqnarray}
\pi_\Phi&=&2b_0\Phi'+b_2(\mathcal{B}'-\lambda),
\\
\pi_\mathcal{B}&=&2b_1\left(\partial^2\Phi+\mathcal{B}'-\lambda\right)+b_2\Phi'+b_3\Phi,
\\
\pi_\lambda &=&0,
\end{eqnarray}
and the Hamiltonian is
\begin{eqnarray}
H&=&\int\D^3x \biggl[\pi_{\cal B}\lambda+
\frac{3}{4\beta}\left(\pi_{\cal B}-\frac{1}{3}\frac{\pi_{\Phi}}{{\cal H}}-b_3\Phi-2b_1\partial^2\Phi \right)^2
+\frac{\pi_{\Phi}^2}{24{\cal H}^2{\cal I}}
-\frac{2{\cal I}+\beta }{3} (\partial^2\Phi)^2
\nonumber\\&& \qquad\quad
 -c_1\Phi^2-c_2\Phi\partial^2\Phi-c_3\Phi{\cal B}-c_4{\cal B}^2-c_5{\cal B}\partial^2\Phi
\biggr].\label{Ham1}
\end{eqnarray}
The primary constraint reads $\pi_\lambda = 0$,
and the consistency of the constraints generates
the following secondary ones:
\begin{eqnarray}
&&
\pi_{\cal B}\approx 0,
\\ &&
c_3\Phi+2c_4{\cal B}+c_5\partial^2\Phi\approx 0,
\\ &&
c_4\lambda+\cdots\approx 0,
\end{eqnarray}
where the last equation fixes $\lambda$. Substituting these constraints to the Hamiltonian~(\ref{Ham1}),
we obtain
\begin{eqnarray}
H_R&=&\int\D^3 x\biggl\{
\frac{2{\cal I}+\beta }{24{\cal H}^2\beta{\cal I}}
\left(\pi_{\Phi}+\frac{6b_3{\cal H}{\cal I}}{2{\cal I}+\beta}
\Phi+4{\cal H}{\cal I}\partial^2\Phi \right)^2
+\left[\frac{c_3^2}{4c_4}-c_1+\frac{3b_3^2}{4(2{\cal I}+\beta)}\right]\Phi^2
\nonumber\\&& \qquad\quad
-\left(c_2-\frac{c_3c_5}{2c_4}-b_3 \right)\Phi\partial^2\Phi
+\frac{c_5^2}{4c_4}(\partial^2\Phi)^2
\biggr\}.
\end{eqnarray}
We thus see that only the two degrees of freedom are left in phase space.
Performing a canonical transfromation,
\begin{eqnarray}
\widetilde \pi_\Phi &=& \pi_\Phi +\frac{6{\cal H I}}{2{\cal I}+\beta}\left(b_3\Phi+2b_1\partial^2\Phi\right),
\\
\widetilde\Phi &=&\Phi,
\end{eqnarray}
whose generating function is given by
\begin{eqnarray}
F=\int\D^3x\left[\widetilde \pi_\Phi\Phi-\frac{3{\cal H}{\cal I}}{2{\cal I}+\beta}
\left(b_3\Phi^2+2b_1\Phi\partial^2\Phi \right)\right],
\end{eqnarray}	
we obtain
\begin{eqnarray}
\widetilde H_R=\int\D^3 x\left[
\frac{2{\cal I}+\beta}{24{\cal H}^2{\cal I}\beta}\widetilde\pi_\Phi^2
+\frac{c_5^2}{4c_4}\left(\partial^2\widetilde\Phi \right)^2 +d_1\widetilde\Phi^2-d_2\widetilde\Phi\partial^2\widetilde\Phi\right],
\end{eqnarray}
where
\begin{eqnarray}
d_1&=&\frac{c_3^2}{4c_4}-c_1+\frac{3b_3^2}{4(2{\cal I}+\beta)}
-3 \left(\frac{b_3{\cal H}{\cal I}}{2{\cal I}+\beta}\right)',
\\
d_2&=&c_2-\frac{c_3c_5}{2c_4}-b_3+2({\cal H}{\cal I})'.
\end{eqnarray}
The stability of the tensor sector has already imposed $\beta >0$.
For a stable kinetic term we therefore require that
\begin{eqnarray}
{\cal I}>0\quad{\rm or}\quad 2{\cal I}+\beta <0.
\end{eqnarray}
Requiring that $c_4>0$, $d_1>0$, and $d_2>0$ is sufficient for $\widetilde H_R$ to be positive definite.
However, one may relax the condition and allow for negative $d_1$ and $d_2$, as
what is crucial is the time scale of instability growth.
In light of this, we have to avoid encountering
the rapid, catastrophic growth of the gradient instability at high momenta,
so that at least we must require $c_4>0$, while $d_1$ and $d_2$ can
be negative for a sufficiently short period.

\section{Primordial Tensor spectrum in Higher Derivative Gravity with Constraints}

The quadratic action for the tensor perturbations with constraints (\ref{2nd-ac-tens-cost})
has a non-standard kinetic term as well as a higher spatial derivative term.
It would therefore be interesting to explore whether or not
this novel structure of the quadratic action gives rise to characteristic imprints
on the primordial tensor spectrum from inflation.
To do so we work in a (quasi-)de Sitter background without assuming
any particular form of the function $f(R,R_{\mu\nu}^2, C_{\mu\nu\rho\sigma}^2)$.
During inflation we have ${\cal H}/a \simeq$ const and ${\cal H}'\simeq {\cal H}^2$,
which leads to $\beta \simeq$ const and
\begin{eqnarray}
A\simeq C\simeq 2f_1+16 f_2\frac{{\cal H}^2}{a^2}\simeq  {\rm const.}
\end{eqnarray}
Thus, ignoring slow-roll suppressed contributions, it suffices to consider
\begin{eqnarray}
S = \frac{1}{8}\int\D^4x\left[h_{ij}'\left(a^2A-2\beta\partial^2 \right)h_{ij}'
+h_{ij}\left(a^2A\partial^2-3\beta\partial^2\partial^2 \right)h_{ij}\right],
\end{eqnarray}
with $a=(-H\eta)^{-1}$, where $H$ is the Hubble scale during inflation.

Following the standard procedure of quantization, we move to the Fourier space and
introduce the canonically normalized variable
\begin{eqnarray}
v_{ij}(\eta; k)&:=&\frac{1}{2}\left(a^2A+2\beta k^2 \right)^{1/2}h_{ij}(\eta; k).
\end{eqnarray}
The equation of motion is given by
\begin{eqnarray}
v_{ij}''+\omega^2v_{ij}=0,\label{eomv}
\end{eqnarray}
where
\begin{eqnarray}
\omega^2:= \frac{1+3\xi k^2\eta^2}{1+2\xi k^2\eta^2}\left(k^2
-\frac{1}{1+2\xi k^2\eta^2}\frac{2}{\eta^2}\right),
\end{eqnarray}
with
\begin{eqnarray}
\xi:=\frac{\beta H^2}{A}\;(={\rm const}) .
\end{eqnarray}
Since $\omega^2\simeq (3/2)k^2$ as $|k\eta|\to \infty$,
the appropriate initial condition is
\begin{eqnarray}
v_{ij}\simeq \frac{1}{(3/2)^{1/4}k^{1/2}}e^{-i\sqrt{3/2}\,k\eta}.\label{initv}
\end{eqnarray}

The equation of motion (\ref{eomv}) can be written
using $y:=-k\eta$ as
\begin{eqnarray}
\frac{\D^2 v_{ij}}{\D y^2}+\frac{1+3\xi y^2}{1+2\xi y^2}\left[1-\frac{2}{(1+2\xi y^2)y^2}\right] v_{ij}=0.
\label{eomv2}
\end{eqnarray}
The growing solution for $y\ll 1$ is given by $v_{ij}\propto 1/y$. Therefore,
the super-horizon solution to Eq.~(\ref{eomv2}) with the initial condition~(\ref{initv})
is of the form
\begin{eqnarray}
v_{ij}\simeq \frac{1}{\sqrt{2k}} \frac{{\cal C}(\xi)}{y},
\end{eqnarray}
where ${\cal C}(\xi)$ should only be characterized by $\xi$.
The primordial tensor spectrum is then given by
\begin{eqnarray}
{\cal P}_h=\frac{{\cal C}^2}{A}\cdot\frac{2H^2}{\pi^2}.
\end{eqnarray}
The power spectrum is nearly scale-invariant, but the amplitude is
modified from the standard result by the model-dependent factor ${\cal C}^2/A$.

\begin{figure}[tb]
  \begin{center}
    \includegraphics[keepaspectratio=true,height=80mm]{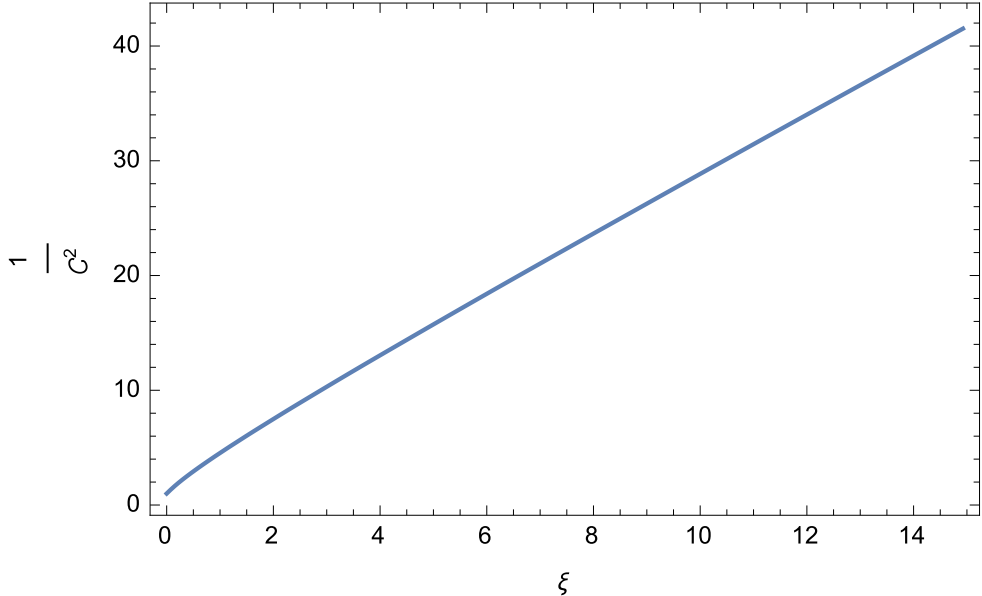}
  \end{center}
  \caption{$1/{\cal C}^2$ as a function of $\xi$.
  }%
  \label{fig:plot_c.eps}
\end{figure}

We numerically solved Eq.~(\ref{eomv2}) for different values of $\xi$ to fix ${\cal C}(\xi)$.
From Fig.~\ref{fig:plot_c.eps} it is found that
\begin{eqnarray}
{\cal C}^2 \simeq \frac{1}{1+s \xi},
\end{eqnarray}
where $s$ is nearly constant even for larger $\xi$ than plotted and $s\gtrsim 2.4$.
Thus, the primordial tensor spectrum is evaluated as
\begin{eqnarray}
{\cal P}_h\simeq \frac{1}{A+s \beta H^2}\frac{2H^2}{\pi^2}.
\end{eqnarray}
It is worth emphasizing that this result is obtained
without assuming any particular form of the function $f(R,R_{\mu\nu}^2, C_{\mu\nu\rho\sigma}^2)$.

\section{Discussion and Conclusions}

In this paper, we have extended the work of Ref.~\cite{Chen:2013aha}
to more general theories of gravity on a less symmetric background,
and shown that cosmological perturbations
in $f(R, R_{\mu\nu}^2, R_{\mu\nu\rho\sigma}^2)$ gravity can be stabilized
by adding constraints to the theory at the level of the quadratic action.
We have found that
the stabilized theory has two tensor and one scalar polarizations.
This indicates that the propagating degrees of freedom are the same as those in
scalar-tensor theories, though it involves safe higher spatial derivatives
of the metric fluctuations such as $\partial_t\partial^2 \delta g_{\mu\nu}$ and $\partial^2\partial^2\delta g_{\mu\nu}$.
It would therefore be intriguing if one could identify the corresponding scalar-tensor theory.
We expect that the ADM description of scalar-tensor theories in the unitary gauge
is the optimal way for this purpose, following and generalizing the
recently developed approach toward single-scalar theories beyond Horndeski~\cite{Gleyzes:2014dya,Gao:2014soa}.
Helpful hints for guessing the corresponding scalar-tensor description
would be obtained by going beyond the quadratic action or by examining
perturbations on more general backgrounds such as Kasner spacetime.
We hope to report our developments in this direction soon.

We have focused on the universal structure of the action for the tensor perturbations
during inflation, and derived the power spectrum of primordial gravitational waves
from the stabilized theory. It would be interesting to evaluate the
power spectrum of the curvature perturbation to confront the stabilized theory with observations.
We will come back to this issue in a future publication.

\acknowledgments
This work was supported in part by JSPS Grant-in-Aid for Young
Scientists (B) No.~24740161 (T.K.). 

\appendix

\section{Coefficients $c_i$ in the quadratic action}\label{appcf}

The following is the list of the coefficients $c_i$ in the
action for the scalar perturbations:
\begin{eqnarray}
c_1&:=&
	-6f_1a^2 \mathcal{H}^2 
	+36 f_{11}\left[4 \mathcal{H}^4-2 \mathcal{H} \mathcal{H}''
	+\left(\mathcal{H}'\right)^2+\mathcal{H}^2 \mathcal{H}'\right]
	+24f_2\left[3 \mathcal{H}^4-2 \mathcal{H} \mathcal{H}''+\left(\mathcal{H}'\right)^2\right]
\nonumber\\&&
	+\frac{72}{a^2} f_{12}
	\left[22 \mathcal{H}^6+4 \left(\mathcal{H}'\right)^3+19 \mathcal{H}^4 \mathcal{H}'-8 \mathcal{H}^3 \mathcal{H}''
	-14 \mathcal{H} \mathcal{H}' \mathcal{H}''+11 \mathcal{H}^2 \left(\mathcal{H}'\right)^2\right]
\nonumber\\&&
	+\frac{216}{a^2}f_{111} \mathcal{H} \left(\mathcal{H}^2+\mathcal{H}'\right) \left(2 \mathcal{H}^3-\mathcal{H}''\right)
\nonumber\\&&
	+\frac{288}{a^4} f_{22}
	\left[12 \mathcal{H}^8+2 \left(\mathcal{H}'\right)^4+21 \mathcal{H}^6 \mathcal{H}'-4 \mathcal{H}^5 \mathcal{H}''
	+11 \mathcal{H}^4 \left(\mathcal{H}'\right)^2
	\right.
\nonumber\\&&\qquad \qquad
	\left.
	-10 \mathcal{H} \left(\mathcal{H}'\right)^2 \mathcal{H}''
	+11 \mathcal{H}^2 \left(\mathcal{H}'\right)^3
	-10 \mathcal{H}^3 \mathcal{H}' \mathcal{H}''\right]
\nonumber\\&&
	+\frac{864}{a^4} f_{112} \mathcal{H} \left[5 \mathcal{H}^7+\mathcal{H} \left(\mathcal{H}'\right)^3
	+7 \mathcal{H}^5 \mathcal{H}'-2 \mathcal{H}^4 \mathcal{H}''+5 \mathcal{H}^3 \left(\mathcal{H}'\right)^2
	-4 \mathcal{H}^2 \mathcal{H}' \mathcal{H}''-3 \left(\mathcal{H}'\right)^2 \mathcal{H}''\right]
\nonumber\\&&
	+\frac{864}{a^6} f_{122} \mathcal{H} \left[16 \mathcal{H}^9+8 \mathcal{H} 
	\left(\mathcal{H}'\right)^4+32 \mathcal{H}^7 \mathcal{H}'-5 \mathcal{H}^6 \mathcal{H}''
	+34 \mathcal{H}^5 \left(\mathcal{H}'\right)^2+18 \mathcal{H}^3 \left(\mathcal{H}'\right)^3
	\right.
\nonumber\\&& \qquad \qquad \quad
	\left.
	-20 \mathcal{H}^2 \left(\mathcal{H}'\right)^2 \mathcal{H}''-12 \left(\mathcal{H}'\right)^3 \mathcal{H}''
	-17 \mathcal{H}^4 \mathcal{H}' \mathcal{H}''\right]
\nonumber\\&&
	+\frac{3456}{a^8} f_{222} \mathcal{H}
	 \left[\mathcal{H}^6+2 \left(\mathcal{H}'\right)^3+3 \mathcal{H}^4 \mathcal{H}'
	+3 \mathcal{H}^2 \left(\mathcal{H}'\right)^2\right] \left[4 \mathcal{H}^5-\mathcal{H}^2 \mathcal{H}''
	+2 \mathcal{H} \left(\mathcal{H}'\right)^2-2 \mathcal{H}' \mathcal{H}''\right],
\nonumber\\&&
\end{eqnarray}
\begin{eqnarray}
c_2&:=&
	2 \left(2 f_2+3 f_{11}\right) \left(4 \mathcal{H}^2+3 \mathcal{H}'\right)
	+\frac{24}{a^2} f_{12} \left[10 \mathcal{H}^4-3 \mathcal{H} \mathcal{H}''
	+6 \left(\mathcal{H}'\right)^2+11 \mathcal{H}^2 \mathcal{H}'\right]
\nonumber\\&&
	+\frac{36}{a^2} f_{111} \mathcal{H} \left(2 \mathcal{H}^3-\mathcal{H}''\right)
\nonumber\\&&
	+\frac{24}{a^4} f_{22} \left[20 \mathcal{H}^6+12 \left(\mathcal{H}'\right)^3
	+35 \mathcal{H}^4 \mathcal{H}'-6 \mathcal{H}^3 \mathcal{H}''
	-12 \mathcal{H} \mathcal{H}' \mathcal{H}''+32 \mathcal{H}^2 \left(\mathcal{H}'\right)^2\right]
\nonumber\\&&
	+\frac{72}{a^4} f_{112} \mathcal{H} \left[8 \mathcal{H}^5-3 \mathcal{H}^2 \mathcal{H}''
	+2 \mathcal{H} \left(\mathcal{H}'\right)^2+8 \mathcal{H}^3 \mathcal{H}'-6 \mathcal{H}' \mathcal{H}''\right]
\nonumber\\&&
	+\frac{144}{a^6} f_{122} \mathcal{H} \left(\mathcal{H}^2+2 \mathcal{H}'\right) 
	\left[10 \mathcal{H}^5-3 \mathcal{H}^2 \mathcal{H}''+4 \mathcal{H} \left(\mathcal{H}'\right)^2
	+4 \mathcal{H}^3 \mathcal{H}'-6 \mathcal{H}' \mathcal{H}''\right]
\nonumber\\&&
	+\frac{288}{a^8} f_{222} \mathcal{H} \left(\mathcal{H}^2+2 \mathcal{H}'\right)^2 \left[4 \mathcal{H}^5
	-\mathcal{H}^2 \mathcal{H}''+2 \mathcal{H} \left(\mathcal{H}'\right)^2-2 \mathcal{H}' \mathcal{H}''\right]~,
\end{eqnarray}
\begin{eqnarray}
c_3&:=&
	-4 f_1 a^2 \mathcal{H}+12 f_{11} \left(8 \mathcal{H}^3-\mathcal{H}''\right)
	+8 f_2 \left(6 \mathcal{H}^3-\mathcal{H}''-2 \mathcal{H} \mathcal{H}'\right)
\nonumber\\&&
	+\frac{48}{a^2} f_{12} \left[26 \mathcal{H}^5-8 \mathcal{H}^2 \mathcal{H}''
	+\mathcal{H} \left(\mathcal{H}'\right)^2+11 \mathcal{H}^3 \mathcal{H}'-2 \mathcal{H}' \mathcal{H}''\right]
	+\frac{216}{a^2} f_{111} \left(2 \mathcal{H}^5-\mathcal{H}^2 \mathcal{H}''\right)
\nonumber\\&&
	+\frac{48}{a^4} f_{22} \left[60 \mathcal{H}^7+4 \mathcal{H} \left(\mathcal{H}'\right)^3+70 \mathcal{H}^5 \mathcal{H}'
	-19 \mathcal{H}^4 \mathcal{H}''
	\right.\nonumber\\&& \qquad \qquad \left.
	+22 \mathcal{H}^3 \left(\mathcal{H}'\right)^2-28 \mathcal{H}^2 \mathcal{H}' \mathcal{H}''
	-4 \left(\mathcal{H}'\right)^2 \mathcal{H}''\right]
\nonumber\\&&
	+\frac{144}{a^4} f_{112} \mathcal{H}^2 \left[28 \mathcal{H}^5-11 \mathcal{H}^2 \mathcal{H}''
	+6 \mathcal{H} \left(\mathcal{H}'\right)^2+20 \mathcal{H}^3 \mathcal{H}'-16 \mathcal{H}' \mathcal{H}''\right]
\nonumber\\&&
	+\frac{288}{a^6} f_{122} \mathcal{H}^2 \left[42 \mathcal{H}^7+20 \mathcal{H} \left(\mathcal{H}'\right)^3
	+68 \mathcal{H}^5 \mathcal{H}'-13 \mathcal{H}^4 \mathcal{H}''+32 \mathcal{H}^3 \left(\mathcal{H}'\right)^2
	\right.
\nonumber\\&&\qquad \qquad \qquad
	\left.
	-40 \mathcal{H}^2 \mathcal{H}' \mathcal{H}''-28 \left(\mathcal{H}'\right)^2 \mathcal{H}''\right]
\nonumber\\&&
	+\frac{576}{a^8} f_{222} \mathcal{H}^2 \left[5 \mathcal{H}^4+8 \left(\mathcal{H}'\right)^2
	+14 \mathcal{H}^2 \mathcal{H}'\right] \left[4 \mathcal{H}^5-\mathcal{H}^2 \mathcal{H}''
	+2 \mathcal{H} \left(\mathcal{H}'\right)^2-2 \mathcal{H}' \mathcal{H}''\right]~,
\end{eqnarray}
\begin{eqnarray}
c_4&:=&
	2 \left(2 f_2+3 f_{11}\right) \left(3 \mathcal{H}^2-\mathcal{H}'\right)
	+\frac{8}{a^2} f_{12} \left[29 \mathcal{H}^4-8 \mathcal{H} \mathcal{H}''
	-5 \left(\mathcal{H}'\right)^2+10 \mathcal{H}^2 \mathcal{H}'\right]
\nonumber\\&&
	+\frac{36}{a^2} f_{111} \mathcal{H} \left(2 \mathcal{H}^3-\mathcal{H}''\right)
\nonumber\\&&
	+\frac{8}{a^4} f_{22} \left[69 \mathcal{H}^6-8 \left(\mathcal{H}'\right)^3+71 \mathcal{H}^4 \mathcal{H}'
	-20 \mathcal{H}^3 \mathcal{H}''-28 \mathcal{H} \mathcal{H}' \mathcal{H}''+18 \mathcal{H}^2 \left(\mathcal{H}'\right)^2\right]
\nonumber\\&&
	+\frac{24}{a^4} f_{112} \mathcal{H} \left[28 \mathcal{H}^5-11 \mathcal{H}^2 \mathcal{H}''
	+6 \mathcal{H} \left(\mathcal{H}'\right)^2+20 \mathcal{H}^3 \mathcal{H}'-16 \mathcal{H}' \mathcal{H}''\right]
\nonumber\\&&
	+\frac{48}{a^6} f_{122} \mathcal{H} \left[42 \mathcal{H}^7+20 \mathcal{H} \left(\mathcal{H}'\right)^3
	+68 \mathcal{H}^5 \mathcal{H}'-13 \mathcal{H}^4 \mathcal{H}''
	\right.\nonumber\\&& \qquad\qquad\quad \left.
	+32 \mathcal{H}^3 \left(\mathcal{H}'\right)^2
	-40 \mathcal{H}^2 \mathcal{H}' \mathcal{H}''-28 \left(\mathcal{H}'\right)^2 \mathcal{H}''\right]
\nonumber\\&&
	+\frac{96}{a^8} f_{222} \mathcal{H} \left[5 \mathcal{H}^4+8 \left(\mathcal{H}'\right)^2+14 \mathcal{H}^2 \mathcal{H}'\right]
	\left[
	4 \mathcal{H}^5-\mathcal{H}^2 \mathcal{H}''+2 \mathcal{H} \left(\mathcal{H}'\right)^2-2 \mathcal{H}' \mathcal{H}''\right].
\end{eqnarray}

\begin{eqnarray}
c_5&:=&4\left(2f_2+3f_{11}\right){\cal H}+\frac{16}{a^2}f_{12}\left(2{\cal H}^3+7{\cal H}{\cal H}'\right)
+\frac{16}{a^4}f_{22}\left[5{\cal H}^5+14{\cal H}^3{\cal H}'+8{\cal H}({\cal H}')^2\right].
\end{eqnarray}



\end{document}